\documentclass[pdftex,twocolumn,epjc3]{svjour3}
\pdfoutput=1

\RequirePackage[T1]{fontenc}

\smartqed  

\RequirePackage{graphicx, multirow}
\RequirePackage{mathptmx, amsmath}      
\RequirePackage{flushend}
\RequirePackage[numbers,sort&compress]{natbib}
\RequirePackage[colorlinks,citecolor=blue,urlcolor=blue,linkcolor=blue]{hyperref}
\usepackage{subcaption}
\usepackage{orcidlink}

\def\bracketbar{\hbox{\kern-9pt\raise1pt%
    \hbox{{\tiny(}{\lower1.5pt\hbox{\bf--}}{\tiny)}}}}

\hypersetup{draft}
\usepackage{lineno}

\journalname{Eur. Phys. J. C}

\begin{document}

\title{A Bayesian approach to the long-baseline neutrino oscillation sensitivity of DUNE}

%

\author{The DUNE Collaboration\\\\
       S.~Abbaslu\thanksref{IPM}
       \and F.~Abd Alrahman\thanksref{Houston}
       \and A.~Abed Abud\thanksref{CERN}
       \and R.~Acciarri\thanksref{CERN}
       \and M.~A.~Acero\thanksref{Atlantico}
       \and M.~R.~Adames\thanksref{Tecnologica }
       \and G.~Adamov\thanksref{Georgian}
       \and M.~Adamowski\thanksref{Fermi}
       \and K.~Adhikari\thanksref{Lancaster}
       \and C.~Adriano\thanksref{Campinas}
% ...
}

\institute{
% ...
\pagebreak[0] Universidad del Atl{\'a}ntico, Barranquilla, Atl{\'a}ntico, Colombia\label{Atlantico}
% ...
\and\pagebreak[0] Universidade Estadual de Campinas, Campinas - SP, 13083-970, Brazil\label{Campinas}
% ...
\and\pagebreak[0] CERN, The European Organization for Nuclear Research, 1211 Meyrin, Switzerland\label{CERN}
% ...
\and\pagebreak[0] Fermi National Accelerator Laboratory, Batavia, IL 60510, USA\label{Fermi}
% ...
\and\pagebreak[0] Georgian Technical University, Tbilisi, Georgia\label{Georgian}
% ...
\and\pagebreak[0] University of Houston, Houston, TX 77204, USA\label{Houston}
% ...
\and\pagebreak[0] Institute for Research in Fundamental Sciences, Tehran, Iran\label{IPM}
% ...
\and\pagebreak[0] Lancaster University, Lancaster LA1 4YB, United Kingdom\label{Lancaster}
% ...
\and\pagebreak[0] Universidade Tecnol{\'o}gica Federal do Paran{\'a}, Curitiba, Brazil\label{Tecnologica }
% ...
}




\onecolumn
\maketitle
\twocolumn
\sloppy

\begin{abstract}

The sensitivity of the Deep Underground Neutrino Experiment (DUNE) to neutrino oscillation is evaluated using a Bayesian Markov Chain Monte Carlo (MCMC) approach. This analysis uses the same underlying sensitivity inputs as previous DUNE studies~\cite{DUNESens}, and therefore does not present updated DUNE sensitivities, but instead explores the additional inferences accessible using a Bayesian approach. We present four-dimensional posterior probability distributions of the oscillation parameters, highlighting the breadth of correlation in the parameter space of interest, especially between $\sin^2 \theta_{23}$ and $\sin^2 \theta_{13}$. We exploit the flexibility of the Bayesian framework to incorporate parameter constraints \textit{post hoc} and assess the impact of applying a reactor short-baseline $\theta_{13}$ constraint. A significant increase in the sensitivity to the $\theta_{23}$ octant is found when including the constraint. Posterior distributions of derived quantities can be easily constructed from MCMC results. This work presents the first study of DUNE's sensitivity to the Jarlskog invariant, $J$, a quantity that provides a parametrisation-independent measure of charge-parity violation in the leptonic sector.

\end{abstract}

\section{Introduction}
\label{sec:intro}

The Deep Underground Neutrino Experiment (DUNE)~\cite{DUNETDR1} is a next-generation, long-baseline neutrino oscillation experiment which will perform a comprehensive study of neutrino mixing using intense $\nu_{\mu}$ and $\bar{\nu}_{\mu}$ beams. The design philosophy of DUNE is to probe the three-flavour framework by measuring all parameters that govern the mixing of $\nu_3$ with $\nu_1$ and $\nu_2$. DUNE’s primary oscillation physics goals are to definitively determine the neutrino mass ordering, measure charge-parity violation (CPV) at high significance if it is present, and precisely measure and test the consistency of the three-flavour neutrino oscillation paradigm. This experimental program will aid theoretical efforts to determine whether symmetries exist in the neutrino sector and whether there is a relationship between the generational structures of quarks and leptons~\cite{leptonquark}. Observation of CPV in neutrinos could be an important step in understanding the origin of the baryon asymmetry of the Universe~\cite{fukugita, davidson}. DUNE's program also includes beyond standard model searches, supernova neutrino detection, and solar neutrino detection. 

DUNE will observe the interactions of neutrinos from an intense neutrino beam with a broad energy distribution with a peak at $\sim$2.5~GeV. The beam will be focused at the near detector (ND), located at Fermi National Accelerator Laboratory in Batavia, Illinois, USA, and the far detector (FD), located at Sanford Underground Research Facility (SURF) in Lead, South Dakota, USA, 1285 km from the neutrino production point. The Long-Baseline Neutrino Facility (LBNF)~\cite{DUNETDR1} will produce the neutrino beam using protons from Fermilab's Main Injector. These protons are incident on a graphite target, and the resulting particles are selected and focused using a horn-focusing system. Depending on the polarity of the focusing magnets, the beam is either $\nu$-enhanced or $\bar{\nu}$-enhanced, corresponding to beams dominated by muon neutrinos and muon antineutrinos, respectively.

In the standard three-flavour neutrino model, neutrino mixing between flavour states $\left\{ \nu_e, \nu_{\mu}, \nu_{\tau} \right\}$ and mass states $\left\{\nu_1, \nu_2, \nu_3\right\}$ is described by the unitary Pontecorvo-Maki-Nakagawa-Sakata (PMNS) matrix~\cite{PDGNeutrinoMixing} and can be parametrized by three mixing angles $\theta_{12}$, $\theta_{13}$ and $\theta_{23}$ and a complex phase $\delta_{CP}$. The magnitude of CP violation is governed by the parametrisation-independent Jarlskog invariant~\cite{Jarlskog}, which, in the parametrisation adopted here, is given by:

\begin{align}
J \equiv & \cos \left(\theta_{12}\right) \sin \left(\theta_{12}\right) \cos \notag
^2\left(\theta_{13}\right) \sin \left(\theta_{13}\right) \\
& \times \cos \left(\theta_{23}\right) \sin \left(\theta_{23}\right) \sin \left(\delta_{\mathrm{CP}}\right)\label{eq:Jarls} .
\end{align}

Neutrino mixing also depends on the two mass-squared splittings $\Delta m^{2}_{21}$ and $\Delta m^{2}_{32}$, where $\Delta m^{2}_{ij} = m^{2}_{i} - m^{2}_{j}$. Current measurements do not yet provide a high-significance determination of the sign of $\Delta m^{2}_{32}$, known as the neutrino mass ordering~\cite{t2knova}, and this remains an important open question.

DUNE is able to probe CPV, the neutrino mass ordering, and make precision measurements of the parameters $\delta_{CP}$, $\theta_{13}$, $\theta_{23}$, and $\Delta m^{2}_{32}$ by measuring the probability that muon (anti)neutrinos maintain their initial flavour simultaneously with the probability that muon (anti)neutrinos oscillate to electron (anti)neutrinos. The latter probability can be written as \cite{neutrinotheory}:

\begin{align}
P\left(\stackrel{\left(-\right)}{\nu}_\mu \rightarrow \stackrel{\left(-\right)}{\nu}_e\right) \simeq \; & \sin ^2 \theta_{23} \sin ^2 2 \theta_{13} \notag\\
& \frac{\sin ^2\left(\Delta_{31}-a L\right)}{\left(\Delta_{31}-a L\right)^2} \Delta_{31}^2 \notag\\
& +\sin 2 \theta_{23} \sin 2 \theta_{13} \sin 2 \theta_{12} \notag\\
& \times \frac{\sin \left(\Delta_{31}-a L\right)}{\left(\Delta_{31}-a L\right)} \Delta_{31} \notag\\
& \times \frac{\sin (a L)}{(a L)} \Delta_{21} \cos \left(\Delta_{31} \pm \delta_{\mathrm{CP}}\right) \notag\\
& +\cos ^2 \theta_{23} \sin ^2 2 \theta_{12} \frac{\sin ^2(a L)}{(a L)^2} \Delta_{21}^2
\end{align}

where

\begin{align}
    a= \pm \frac{G_{\mathrm{F}} N_e}{\sqrt{2}} \approx \pm \frac{1}{3500 \mathrm{~km}}\left(\frac{\rho}{2.848 \mathrm{~g} / \mathrm{cm}^3}\right),
\end{align}

$G_{\mathrm{F}}$ is the Fermi constant, $N_e$ is the number density of electrons in the Earth's crust, $\rho$ is the Earth crust matter density, $\Delta_{ij}$ = $1.267 \Delta m^{2}_{ij} L/E_\nu$, L is the baseline in km, and $E_\nu$ is the neutrino energy in GeV. Both $\delta_{CP}$ and $a$ terms are positive (negative) for $\nu_\mu \rightarrow \nu_e \left( \bar{\nu}_\mu \rightarrow \bar{\nu}_e \right)$ oscillations. Therefore, a neutrino-antineutrino asymmetry is introduced both by CPV, i.e. $\delta_{CP}$, and the matter effect, i.e. $a$. The asymmetry as a result of the matter effect originates from the abundance of electrons and the absence of positrons in the Earth's crust.

 In this work, DUNE’s sensitivity to the neutrino oscillation parameters is studied using a Bayesian statistical treatment. This work follows from previously published classical frequentist sensitivity estimates~\cite{DUNESens}, and uses a more sophisticated systematic uncertainty implementation of the same modelling components. The Bayesian treatment allows for a flexible analysis of the complex, multidimensional parameter space, including the first presentation of DUNE's sensitivity to $J$. 

\section{Efficient Bayesian Inference}
\label{sec:bayes}  

Bayesian inference determines the posterior probability distribution of the model parameters given the observed data and prior information. The posterior is influenced by the likelihood, i.e. the parametrised probability of observing the data, and the prior probabilities on each parameter of that likelihood. Direct, analytical evaluation of the posterior probability is often intractable, so we use Markov Chain Monte Carlo (MCMC) sampling to extract the posterior, specifically the Metropolis--Rosenbluth--Rosenbluth--Teller--Teller (MR\textsuperscript{2}T\textsuperscript{2}) algorithm~\cite{mr2t2}.

This algorithm requires a proposal function to tell it how to move around parameter space, with a multivariate Gaussian being a common choice. However, the covariance chosen for this Gaussian has a large impact on the sampling efficiency and manually tuning it with hundreds of parameters is a time consuming process. 

Since the optimal proposal function is a Gaussian approximation of the posterior that is being sampled, this problem lends itself to an iterative solution, known as the Adaptive Metropolis (AM) algorithm~\cite{amcmc}. AM iteratively updates the proposal covariance by calculating the covariance of the steps already taken in the chain. We perform this process at the beginning of our MCMC sampling and find the tuned covariance requires $\sim 50\text{x}$ fewer MCMC steps to sufficiently sample the space compared to a manually chosen covariance.

Furthermore, the computational cost of AM can be reduced by performing the iterative tuning part of the fit on a small random subset of Monte Carlo (MC) events upweighted to produce the same predicted data rate. This process is called downsampling. Once a sufficiently efficient covariance is obtained we discard the steps already taken and revert to using the full MC sample for the final fit. In this work, AM was performed on a 5\% subset of the MC events, resulting in an $\sim 15\text{x}$ reduction in the computational cost in optimising the proposal function.

\subsection{Prior Reweighting}
\label{sec:reweight}
In the case where there are multiple justified choices of priors, the effect on the posterior of switching between these choices should be understood. A further advantage of MCMC is that the effect on the posterior from changing between priors is easy to calculate. Rather than recomputing the MCMC samples under the new prior, each MCMC sample can be reweighted by the ratio between the new and old priors~\cite{PriorReweight}.

For this process to be accurate, it is necessary for the regions preferred by the new prior to have been sampled sufficiently under the old prior for there to be sufficient steps to weight up. This technique can also be used to present the impact of new experimental measurements or constraints on important model parameters.

\section{Analysis Framework}
\label{sec:analysisframework}

Predicting the neutrino event spectrum that DUNE will observe requires sufficient modelling of the neutrino flux produced by the beam, neutrino interactions and the detectors. This section will provide an overview of the models and the implementation of their associated uncertainties. A detailed description of the models can be found in~\cite{DUNESens}. The oscillation analyses presented in this work use the MaCh3 framework~\cite{MaCh3}.

\subsection{Neutrino Flux}
\label{sec:flux}

The LBNF beam will operate in two modes, a neutrino-enhanced and an antineutrino-enhanced mode. Each mode is dominated by muon-flavour (anti)neutrinos. Both modes will produce a neutrino beam with a peak energy of $\sim 2.5$~GeV. The neutrino flux prediction is generated using G4LBNF~\cite{DUNETDR2, numi}, a simulation of the LBNF beamline. The flux uncertainties are driven by the modelling of high-energy hadron interactions in the target and by beamline component design tolerances; e.g. the position and orientation of the focusing horns. The estimated total flux uncertainty is $\sim 8\%$ at the first oscillation maximum and $\sim 12\%$ at the second maximum. These uncertainties are highly correlated between beam modes, neutrino flavours, and energy bins.  

\subsection{Neutrino Interactions}
\label{sec:xsec}

The neutrino interaction model used in this analysis is based on version 2.12.10 of the GENIE event generator~\cite{GENIE, GENIEManual}. The uncertainty on neutrino interactions accounts for free theory parameters in the GENIE model and additional \textit{ad hoc} uncertainties developed to expand the GENIE error envelope to better cover current-generation neutrino scattering measurements and some alternate interaction models. A detailed description of the model is given in Ref.~\cite{DUNESens}.

In this analysis, an improved implementation of the approximate response functions used to modify the predicted event rate for variations of the interaction model parameters is used. A separate approximate response function is constructed for each observable bin and simulated neutrino interaction process. This reduces the impact of assuming perfect factorisability of the interaction cross section over all free parameters.

\subsection{Detector and Event Classification Uncertainties}
\label{sec:det}

The energy of the incoming neutrino in CC events is estimated as the sum of the lepton and hadronic energies reconstructed using the Pandora framework~\cite{PandoraSoft, PandoraDUNE}. Uncertainty in the calibrated total and per-particle energy scales and per-particle resolutions, as well as the acceptance of the detector, is parametrised as in  Ref.~\cite{DUNESens}. It might be expected that the detector response models for the near and far detector should be correlated. As in Ref.~\cite{DUNESens}, the conservative approach of leaving these effects uncorrelated is taken.

As in Ref.~\cite{DUNESens}, the ND detector uncertainties are encoded into a covariance matrix and left unconstrained in the ND analysis. This is done to protect against over-constraint that could be caused by the parameterised ND reconstruction. The uncertainty treatment at the FD is improved over the previous analysis. Systematic detector variations in the FD can now cause events to migrate between observable bins, further reducing the impact of assumed factorisability in the systematic model.

Event classification at the FD is performed using a convolutional neural network (CNN). A detailed description of the architecture can be found in Ref.~\cite{CVN}. The CNN assigned a confidence score for each of the following sample labels: Charged-Current Inclusive (CCInc) $\nu_\mu$ and CCInc $\nu_e$ in neutrino-enhanced beam mode, as well as CCInc $\bar\nu_\mu$ and CCInc $\bar\nu_e$ in antineutrino-enhanced mode. Analysis samples are selected by cutting on the label scores. An \textit{ad hoc} 1\% uncertainty is applied to the score value of each event as in Ref.~\cite{DUNESens}. Since comprehensive reconstruction had not been finalised for the ND at the time of this analysis, event classification at the ND is performed using track length and mean energy deposits to distinguish between pions and muons as described in Ref.~\cite{DUNESens}.

Although these improvements have only a limited impact on the sensitivity studies presented here, they provide a more realistic description of systematic effects and their propagation through the oscillation analysis. This capability will become increasingly important as DUNE moves from sensitivity studies to the interpretation of real data, where accurate modelling of systematic uncertainties is essential for robust parameter estimation.

\section{Sensitivity Methods}
\label{sec:SensMethod}

Previous DUNE sensitivity studies used the CAFAna framework~\cite{cafana}. This sensitivity study uses the MaCh3 framework~\cite{MaCh3} to perform the analysis. Systematics are implemented as described in \autoref{sec:analysisframework}, resulting in a comprehensive model that combines binned response functions, normalisation parameters, and event migration. The systematic uncertainty treatment and resulting oscillation parameter inference were cross-validated between the two frameworks and found to be in good agreement. The oscillation probability for a given hypothesis is calculated in fine (10 MeV) true neutrino energy bins using the CUDAProb3 framework~\cite{CUDAProb3}. Thus, a prediction of the event distribution can be produced both at the FD and the ND, given parameter inputs for the flux, cross-section, detector models and oscillation parameter values. The multi-dimensional likelihood space, resulting from the comparison between the simulated prediction and simulated data spectrum, is explored using MCMC (as described in \autoref{sec:bayes}). 

Oscillation parameters of interest, i.e. $\sin^2 \theta_{23}$, $\sin^2 \theta_{13}$, $\Delta m^2_{32}$ and $\delta_{CP}$, are included in the fit with uniform priors. This is standard practice for physics parameters we intend to measure directly. Since short-baseline reactor antineutrino experiments have precisely measured $\theta_{13}$, where stated, an external Gaussian constraint is applied to $\sin^2\theta_{13}$ via prior reweighting (described in \autoref{sec:reweight}). A Gaussian prior is also applied to $\sin^2 \theta_{12}$ and $\Delta m^2_{21}$, since it is not expected that DUNE's beam analysis will have significant sensitivity to these parameters\footnote{The inclusion of solar samples would provide sensitivity to these parameters and is an area of active development within the experiment.}.

The flux, cross-section and FD detector parameters are included with Gaussian priors corresponding to the pre-fit uncertainty. The fit to the ND samples constrains these parameters, and the resulting full posterior distribution is approximated by a multivariate Gaussian. This Gaussian is then used to define the pre-fit uncertainties and correlations of the systematic parameters in the FD fit. As explained in \autoref{sec:det}, the ND detector uncertainty is not constrained by the ND samples. The decision to perform the ND fit separately and include it as an external constraint is driven by computational considerations. The ND MC dataset contains many more simulated neutrino interactions than the FD, requiring more computational resources to perform a fit. Separating the two allows a single ND fit to be used on multiple simulated FD datasets with different true oscillation parameter hypotheses. Negligible differences are observed between the posterior distributions and credible intervals obtained using the external constraint and those obtained from a simultaneous fit to the FD and ND samples, validating the use of the external constraint in this work.

The posterior, which indicates the compatibility of a given oscillation hypothesis with the data while incorporating prior information, is often conveniently expressed in the negative logarithmic space, which results in the following expression:

\begin{align}
-\log \mathcal{L}_{\mathrm{Total}}
={}&
\frac{1}{2}
\sum_{k=1}^{N_{\mathrm{params}}}
\sum_{l=1}^{N_{\mathrm{params}}}
(\theta_k-\mu_k)
V^{-1}_{kl}
(\theta_l-\mu_l)
\notag\\[4pt]
&+
\begin{cases}
\displaystyle
\sum_{i=1}^{N_{\mathrm{bins}}}
\left[
\lambda_i-n_i
+n_i\log\!\left(\dfrac{n_i}{\lambda_i}\right)
\right]
& \mathrm{(FD)}
\\[12pt]
\displaystyle
\sum_{i,j=1}^{N_{\mathrm{bins}}}
(n_i-\lambda_i)
C^{-1}_{ij}
(n_j-\lambda_j)
& \mathrm{(ND)}
\end{cases}
\label{eq:loglikelihood}
\end{align}
where $\vec{\theta}$ is the vector of both oscillation and nuisance parameters; $\mu_{k}$ is the central value of the $k$th model parameter; $\textbf{V}_{kl}^{-1}$ is the prior covariance matrix which holds all prior uncertainties and correlations of the model parameters and $\lambda_{i}$ and $n_{i}$ are the MC prediction and the simulated data in the $i$th bin respectively.  For the ND fit, the Poisson likelihood term is replaced with a Gaussian penalty term, where $\text{C}_{ij}^{-1}$ is the covariance matrix mentioned in \autoref{sec:det}, produced by "throwing" the ND detector parameters.

The results presented in this work are based on Asimov studies~\cite{asimov}. These are studies in which the simulated (Asimov) dataset is generated from the nominal MC. A range of Asimov datasets are used in this work, corresponding to different true values of oscillation parameters and different exposures. The values of $\sin^2 \theta_{23}$, $\sin^2 \theta_{13}$ and $\Delta m^{2}_{32}$ used to produce the Asimov dataset are taken from two previous results; NuFIT 6.0 global fit (with Super-K atmospheric data)~\cite{nufit6}, shown in \autoref{tab:NuFit6}, and the recent T2K-NOvA joint analysis best-fit point in inverted ordering~\cite{t2knova}, shown in \autoref{tab:T2KNOvA}. These two results are chosen since they each represent different relevant parts of the oscillation parameter space. The NuFIT 6.0 best-fit lies in the normal ordering (NO) and the lower $\theta_{23}$ octant, while the T2K-NOvA best fit lies in the inverted ordering (IO) and the upper octant. For both NuFIT 6.0 and T2K–NOvA Asimov datasets, the mass ordering sensitivity is not considered, since at these exposures DUNE is expected to resolve the mass ordering at >$5\sigma$, irrespective of the true values of the other oscillation parameters~\cite{DUNETDR2}. The best-fit values of $\delta_{CP}$ in both of these Asimov points are replaced with $\delta_{CP} = 0, -\frac{\pi}{4}, -\frac{\pi}{2}$. The motivation behind this is, given that the value of $\delta_{CP}$ is currently not well measured, the sensitivity of DUNE when $\delta_{CP}$ is minimal, maximal, and somewhere in between can be assessed.

\begin{table}[]
\centering
\renewcommand*{\arraystretch}{1.3}
\begin{tabular*}{\columnwidth}{@{\extracolsep{\fill}}lcc}
\hline
\textbf{Parameter} & \textbf{Central Value} & \textbf{Uncertainty} \\
\hline
$\sin^2\theta_{12}$ & 0.308 & 0.012 \\
$\sin^2\theta_{23}$ & 0.470 & uniform prior \\
$\sin^2\theta_{13}$ & 0.02215 & uniform prior \\
\hline
$\Delta m^{2}_{21}\,(\text{eV}^2)$ & $7.49 \times 10^{-5}$ & $1.9 \times 10^{-6}$ \\
$\Delta m^{2}_{32}\,(\text{eV}^2)$ & $2.438 \times 10^{-3}$ & uniform prior \\
\hline
$\delta_{CP}\,(\mathrm{rad})$ & $0,\ -\pi/4,\ -\pi/2$ & uniform prior \\
\hline
\multicolumn{3}{c}{\textit{Mass Ordering: Normal}} \\
\hline
\end{tabular*}
\caption{Central values and uncertainties of the neutrino oscillation parameters from the NuFIT 6.0 global fit (including SK atmospheric data)~\cite{nufit6}. The values of $\delta_{CP}$ are modified from the best-fit values to those used in this study.}
\label{tab:NuFit6}
\end{table}

\begin{table}[]
\centering
\renewcommand*{\arraystretch}{1.3}
\begin{tabular*}{\columnwidth}{@{\extracolsep{\fill}}lcc}
\hline
\textbf{Parameter} & \textbf{Central Value} & \textbf{Uncertainty} \\
\hline
$\sin^2\theta_{12}$ & 0.308 & 0.012 \\
$\sin^2\theta_{23}$ & 0.563 & uniform prior \\
$\sin^2\theta_{13}$ & 0.02196 & uniform prior \\
\hline
$\Delta m^{2}_{21}\,(\text{eV}^2)$ & $7.49 \times 10^{-5}$ & $1.9 \times 10^{-6}$ \\
$\Delta m^{2}_{32}\,(\text{eV}^2)$ & $-2.43 \times 10^{-3}$ & uniform prior \\
\hline
$\delta_{CP}\,(\mathrm{rad})$ & $0,\ -\pi/4,\ -\pi/2$ & uniform prior \\
\hline
\multicolumn{3}{c}{\textit{Mass Ordering: Inverted}} \\
\hline
\end{tabular*}
\caption{Central values and uncertainties of the neutrino oscillation parameters from the T2K--NOvA joint analysis best-fit~\cite{t2knova}. The values of $\delta_{CP}$ are modified from the best-fit values to those used in this study.
}
\label{tab:T2KNOvA}
\end{table}







To improve the efficiency of the MCMC method in sampling the likelihood space, different proposal functions are chosen depending on the parameter and the Asimov dataset being studied. Since systematic parameters generally have simple Gaussian distributions, a Gaussian proposal function is sufficient for exploring the parameter space effectively. This is also true for $\sin^2 \theta_{12}$ and $\Delta m^{2}_{21}$, where the imposed prior ensured a Gaussian distribution. Since there exists a degeneracy in the upper and lower octants of $\theta_{23}$, a "flip" is introduced into the MCMC proposal function for $\sin^2 \theta_{23}$, on top of the standard Gaussian proposal, with a 50\% probability of occurring. This allows the exploration of both octants without the need for the chain to "cross" the maximal mixing ($\sin^2 \theta_{23}$ = 0.5) region in between, which is of low likelihood. The octant of $\theta_{23}$ is highly correlated with $\sin^2 \theta_{13}$, resulting in two well-defined lobes in the 2D parameter space. Therefore, an octant flip in just $\sin^2 \theta_{23}$ is inefficient unless also applied to $\sin^2 \theta_{13}$. This approach shows good sampling efficiency for both upper and lower octant Asimov datasets. For $\delta_{CP}$, a flip is applied about $\delta = \pm \pi/2$, since large degeneracies can be present, especially for Asimov datasets where $\delta_{CP} = 0$. Since the sensitivity to the mass ordering is not evaluated in this study, it is unnecessary to impose a flip on the sign of $\Delta m^2_{32}$. A Gaussian proposal function is sufficient for sampling the distribution of $\Delta m^2_{32}$ in the correct ordering.

As well as changing the true value of oscillation parameters, two different FD exposures of 200~kt-MW-yr and 1000~kt-MW-yr are used for each of the Asimov datasets described. Throughout this work, exposure is quoted in units of kt-MW-yr, defined as the product of the fiducial detector mass (kilotonnes), beam power (megawatts), and running time (years). The 200~kt-MW-yr exposure represents a benchmark for what can be achieved with 2 FD modules (20 kt) and a 1.2 MW beam. The 1000 kt-MW-yr exposure represents the long-term reach of the experiment. This results in a total of 12 Asimov datasets used in this work. 

\begin{figure*}[]
\centering

\begin{subfigure}{0.48\textwidth}
  \centering
  \includegraphics[width=\textwidth]{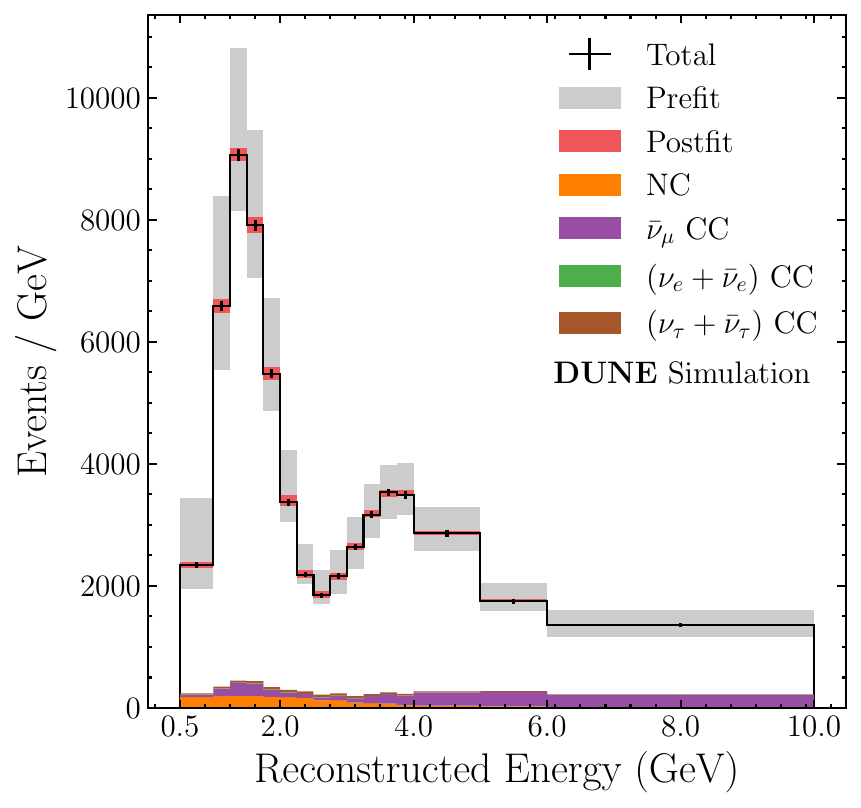}
  \caption{}
  \label{fig:FDFHCNumuPostPred}
\end{subfigure}
\hfill
\begin{subfigure}{0.48\textwidth}
  \centering
  \includegraphics[width=\textwidth]{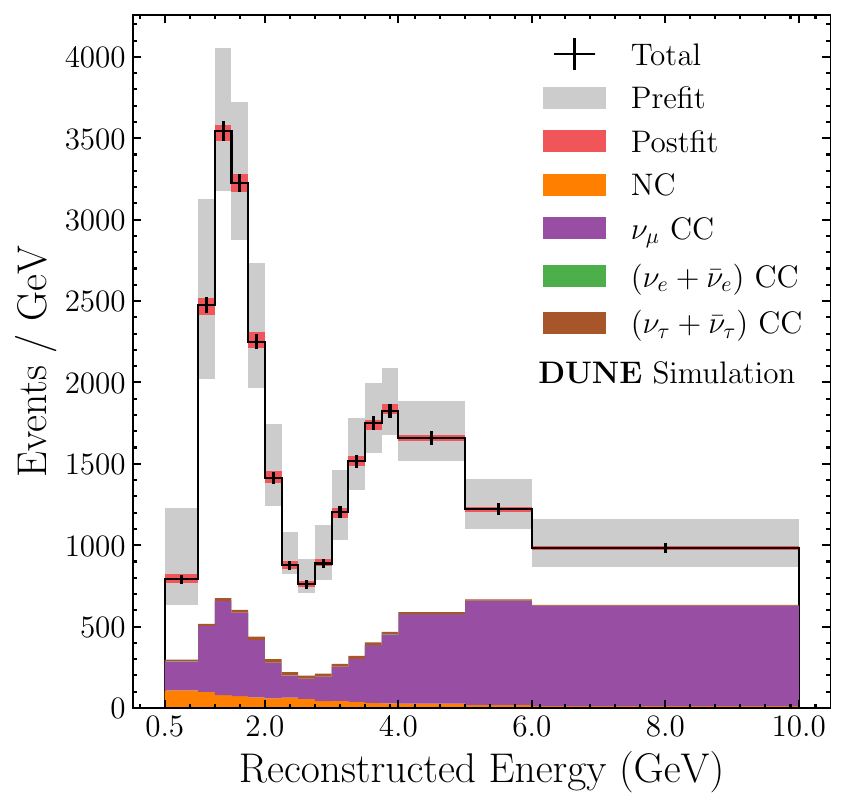}
  \caption{}
  \label{fig:FDRHCNumuPostPred}
\end{subfigure}

\vspace{0.3cm}

\begin{subfigure}{0.48\textwidth}
  \centering
  \includegraphics[width=\textwidth]{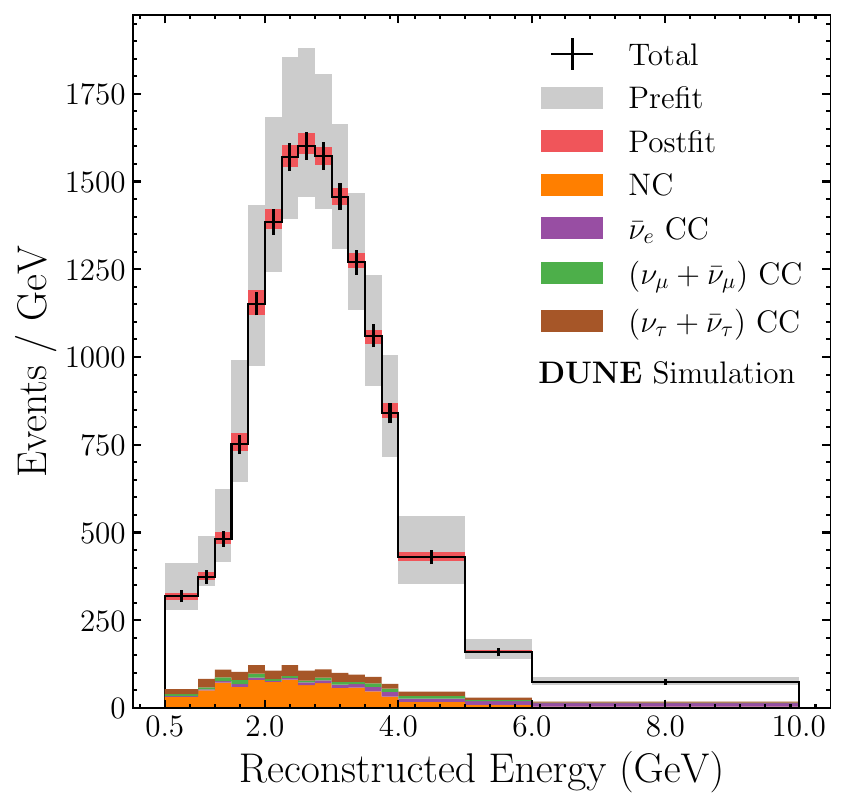}
  \caption{}
  \label{fig:FDFHCNuePostPred}
\end{subfigure}
\hfill
\begin{subfigure}{0.48\textwidth}
  \centering
  \includegraphics[width=\textwidth]{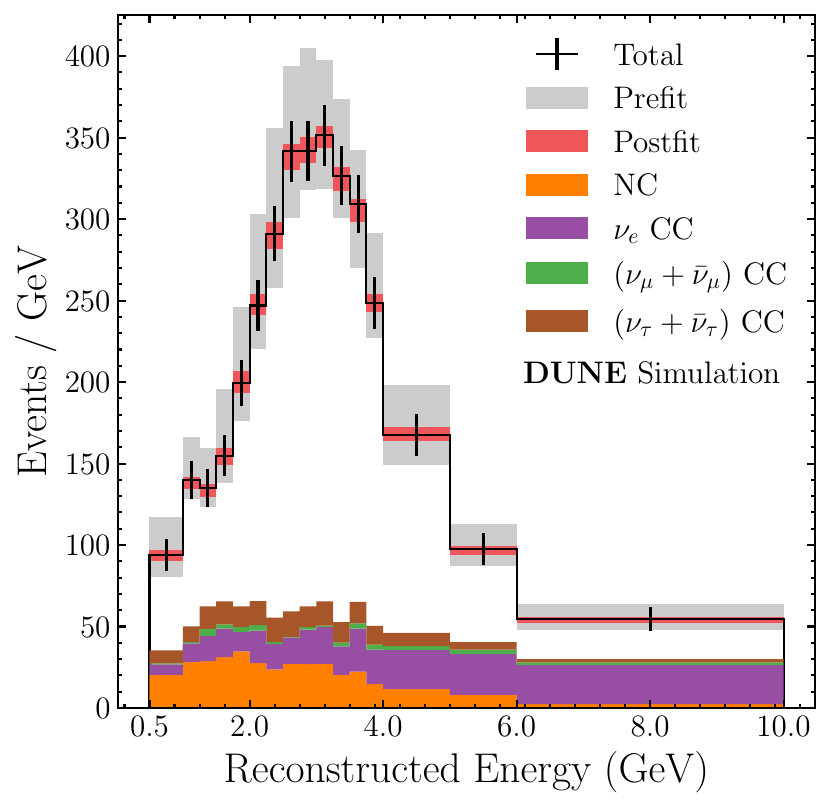}
  \caption{}
  \label{fig:FDRHCNuePostPred}
\end{subfigure}

\caption{
Reconstructed neutrino energy distributions of the (a) $\nu$-enhanced $\nu_{\mu}$-like, (b) $\bar{\nu}$-enhanced $\bar{\nu}_{\mu}$-like, (c) $\nu$-enhanced $\nu_{e}$-like, and (d) $\bar{\nu}$-enhanced $\bar{\nu}_{e}$-like FD samples for a 1000~kt-MW-yr exposure with an equal split between $\nu$-enhanced and $\bar{\nu}$-enhanced run modes. The distributions are produced with oscillation parameters set to the NuFIT 6.0 best-fit values (see~\autoref{tab:NuFit6}). The size of the systematic uncertainty band from all flux, cross-section, and FD detector systematic uncertainties used in the analysis is shown, together with the posterior uncertainty bands obtained by performing an Asimov fit to the FD data. NC backgrounds and wrong-sign contributions to the total event rate are also shown.
}
\label{fig:FDPostPred}

\end{figure*}



The oscillation analysis presented here includes two ND samples, reconstructed $\nu_\mu$ candidates from the $\nu$-enhanced beam mode and $\bar{\nu}_\mu$ candidates from the $\bar{\nu}$-enhanced beam mode, of interactions on a detector with a similar technology as the FD. Both samples are CC-inclusive, referring to charged-current interactions regardless of final-state particles produced. These samples are binned in two dimensions, as a function of reconstructed neutrino energy and reconstructed inelasticity, $\text{y}_{\text{rec}} = 1 - \text{E}_{\mu}^{\text{rec}}/\text{E}_{\nu}^{\text{rec}}$. The very high number of events at the ND results in a significantly constrained systematic uncertainty. However, since the statistical uncertainty in each bin is much smaller than the systematic uncertainty, this suggests the ND samples are systematics limited with the binning used in this analysis. In future analyses, more sophisticated samples and binning choices will allow more to be extracted from the ND data, resulting in better mitigation of systematic uncertainty. For example, the inclusion of neutrino-electron scattering samples is expected to independently constrain the neutrino flux normalisation to approximately 2\%~\cite{NueScatter}.

\begin{figure*}[t]
    \centering
    \includegraphics[width = \textwidth]{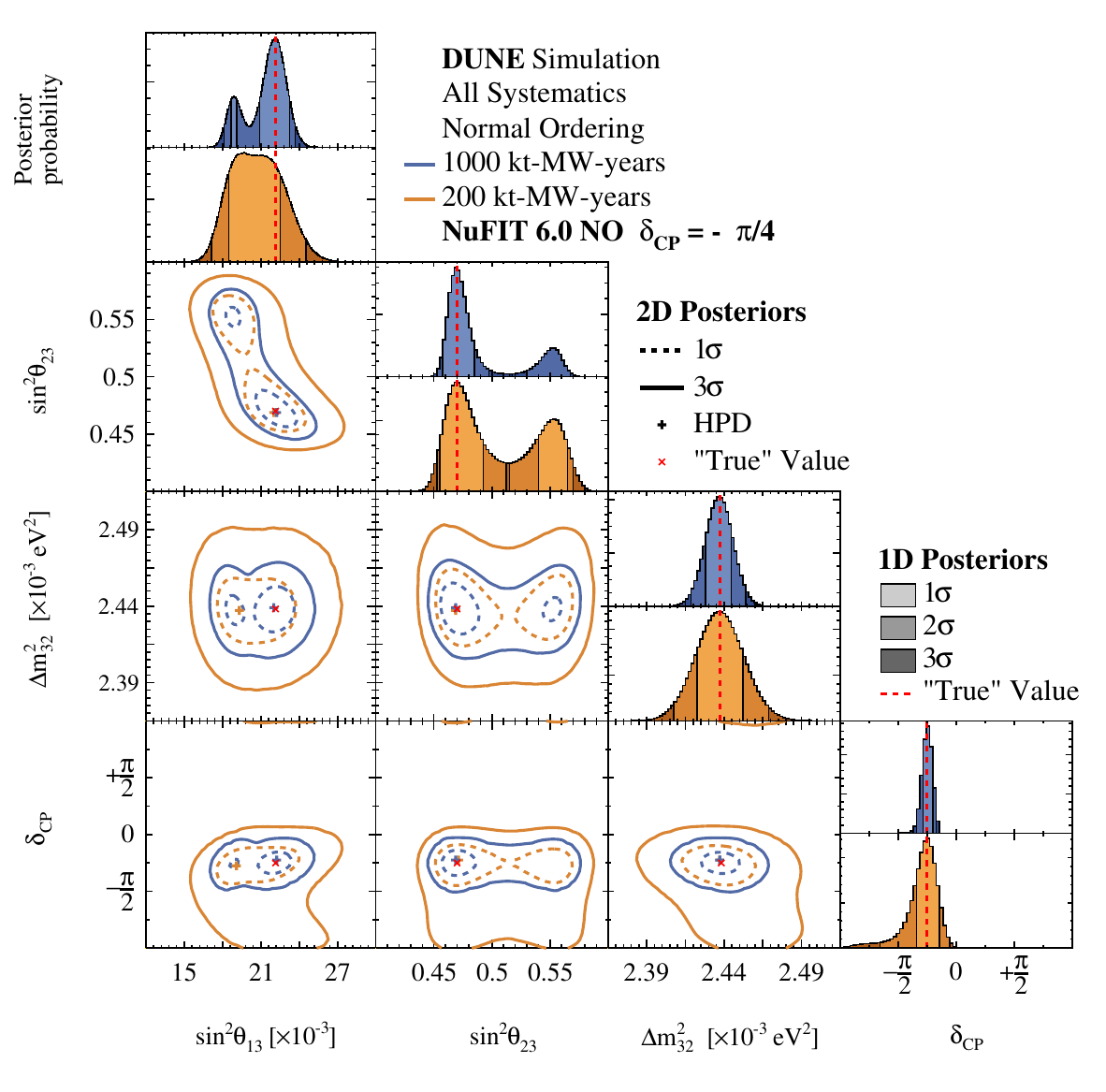}
    \caption{Comparison of the four-dimensional posterior distribution in the mixing angles and $\Delta m^{2}_{32}$ between 200~kt-MW-yr (gold) and 1000~kt-MW-yr (blue) exposures at NuFIT 6.0 $\delta_{CP} = -\pi/4$ Asimov point. The 2D projections compare the 1$\sigma$ and 3$\sigma$ credible intervals. The 1D projections compare the 1D posterior as well as 1, 2 and 3 $\sigma$ intervals. The underlying 2D posteriors are omitted to ease comparison between exposures.}
    \label{fig:NuFit6CompDcpMpi4}
\end{figure*}

Four CC-inclusive reconstructed FD samples are included in the analysis: $\nu_\mu$ candidates from the $\nu$-enhanced beam mode, $\nu_e$ candidates from the $\nu$-enhanced beam mode, $\bar{\nu}_\mu$ candidates from the $\bar{\nu}$-enhanced beam mode and $\bar{\nu}_e$ candidates from the $\bar{\nu}$-enhanced beam mode, and are binned in reconstructed neutrino energy. These samples provide sensitivity to the oscillation parameters. The $\nu$-enhanced $\nu_\mu$ and $\bar{\nu}$-enhanced $\bar{\nu}_\mu$ samples, known as the disappearance spectra, provide sensitivity to $\sin^2 2\theta_{23}$, which affects the depth of the oscillation dip in the spectrum, as well as $\Delta m^{2}_{32}$, which controls the energy at which the oscillation dip occurs. Since the disappearance probability is a function of $\sin^2 2\theta_{23}$, there is degeneracy between the upper octant, $\sin^2 \theta_{23} > 0.5$, and the lower octant, $\sin^2 \theta_{23} < 0.5$, of $\theta_{23}$. The  $\nu_e$ and $\bar{\nu}$-enhanced $\bar{\nu}_e$ samples, known as the appearance spectra, provide sensitivity to $\sin^2 \theta_{13}$ and $\sin^2 \theta_{23}$, where the dependence on $\sin^2 \theta_{23}$ provides sensitivity to the octant. Sensitivity to $\delta_{CP}$ arises from the rate and shape of the appearance spectra, as well as the differences in the $\nu_e$ and $\bar{\nu}_e$ samples.

The FD $\nu$-enhanced $\nu_\mu$ and $\bar{\nu}$-enhanced $\bar{\nu}_\mu$ samples are shown in \autoref{fig:FDFHCNumuPostPred} and \autoref{fig:FDRHCNumuPostPred} for a 1000~kt-MW-yr exposure, split equally between $\nu$-enhanced and $\bar{\nu}$-enhanced beam modes. The pre-fit systematic uncertainty envelope is shown, along with the envelope after an FD Asimov fit using the ND constraint, although the systematic uncertainties are already almost entirely constrained by the ND samples. Only events with reconstructed neutrino energies between 0.5~GeV and 10~GeV are included in all FD samples. The prior systematic uncertainty is largest around the peak of both distributions. At 1000~kt-MW-yr exposure, the statistical uncertainty is much smaller than the systematic uncertainty band after the ND constraint, even in the "dip" region at 2.5~GeV. This region is especially important as it is most sensitive to the disappearance parameters. The $\bar{\nu}$-enhanced $\bar{\nu}_\mu$ sample has significantly fewer events than the $\nu$-enhanced $\nu_\mu$ sample, owing to both the lower flux in $\bar{\nu}$-enhanced mode and the smaller $\bar{\nu}_\mu$ cross-section. Although this results in a larger statistical uncertainty, the post-fit systematic uncertainty remains dominant. Despite the lower $\bar{\nu}_\mu$ event rate, studies have shown that equal time spent in each run mode provides optimal sensitivity to the mass ordering and CPV \cite{LowExposure}. The background contributions to the samples are also shown. Wrong-sign neutrinos are the largest background in both samples, but are significantly larger in the $\bar{\nu}$-enhanced sample.

The FD $\nu$-enhanced $\nu_e$ and $\bar{\nu}$-enhanced $\bar{\nu}_e$ samples are shown in \autoref{fig:FDFHCNuePostPred} and \autoref{fig:FDRHCNuePostPred} for a 1000~kt-MW-yr exposure, split equally between $\nu$-enhanced and $\bar{\nu}$-enhanced beam modes. The pre-fit systematic uncertainty envelope is shown, as well as the envelope after an FD Asimov fit using the ND constraint. There are far fewer events than in \autoref{fig:FDFHCNumuPostPred} and \autoref{fig:FDRHCNumuPostPred}; the statistical uncertainty is comparable to the systematic uncertainty after the ND constraint is applied.

\section{Sensitivities}

In this section, various sensitivity results are presented, all of which are inferred from the full posterior probability distribution for a given Asimov dataset. Information on just the oscillation parameters of interest is extracted via marginalisation. This requires integrating over all nuisance parameters (including $\sin^2 \theta_{12}$ and $\Delta m^2_{21}$). The highest posterior density point (HPD) indicates the bin with the largest number of MCMC samples. It is not always the case that the HPD should line up with the true Asimov point, since marginalisation effects can cause the HPD in one or two dimensions to deviate from the HPD in the full parameter space. The credible intervals presented in this work correspond to the regions of parameter space which contain 68.3\% (1$\sigma$), 95.4\% (2$\sigma$) and 99.7\% (3$\sigma$) of the posterior distribution. 

\autoref{fig:NuFit6CompDcpMpi4} shows the 4D posterior for both 200~kt-MW-yr and 1000~kt-MW-yr exposures at the NuFIT 6.0 $\delta_{CP} = -\pi/4$ Asimov point. This presentation of parameter measurements highlights the power of the Bayesian approach by showing the breadth of correlations in the 4-dimensional oscillation space to which DUNE is sensitive. The strong anti-correlation between $\sin^2\theta_{23}$ and $\sin^2\theta_{13}$ is evident in the relevant 2D posterior. 

At 200~kt-MW-yr exposure, there is little preference for the octant of $\theta_{23}$. Since the Asimov point is in the lower octant, the lower $\nu_{e}$ event rate has a significant effect on sensitivity. This prevents even regions around maximal mixing from being disfavoured significantly. Another interesting feature is the $\sin^2\theta_{13}$ posterior. The HPD of the 1D posterior does not align with the true value. This is a marginalisation effect caused by the strong degeneracy between $\sin^2\theta_{13}$ and $\sin^2\theta_{23}$. Although the true value remains within the 1$\sigma$ credible interval, this demonstrates the importance of considering the full multi-dimensional posterior distribution, as done in \autoref{fig:NuFit6CompDcpMpi4}. In the corresponding $\sin^2\theta_{23} - \sin^2\theta_{13}$ 2D posterior, the correlation is resolved and the HPD aligns with the true value. Weak correlations can be seen in the 2D 3$\sigma$ credible intervals of $\delta_{CP}$ with all other oscillation parameters of interest. There is a significant tail in the $\delta_{CP}$ distribution towards negative values.

At 1000~kt-MW-yr, the degeneracies across the entire 4D parameter space are lifted significantly. There is a reduction in the correlations between $\delta_{CP}$ and the remaining oscillation parameters. The tail in the distribution of $\delta_{CP}$ is no longer present. At this increased exposure, maximal mixing no longer lies within the 1$\sigma$ credible interval, and there is a stronger preference for the true octant. This, in turn, mitigates the marginalisation effect in the 1D $\sin^2\theta_{13}$ posterior, resulting in agreement between the HPD and the true value.

\begin{figure*}[t]
    \centering
    \includegraphics[width = \textwidth]{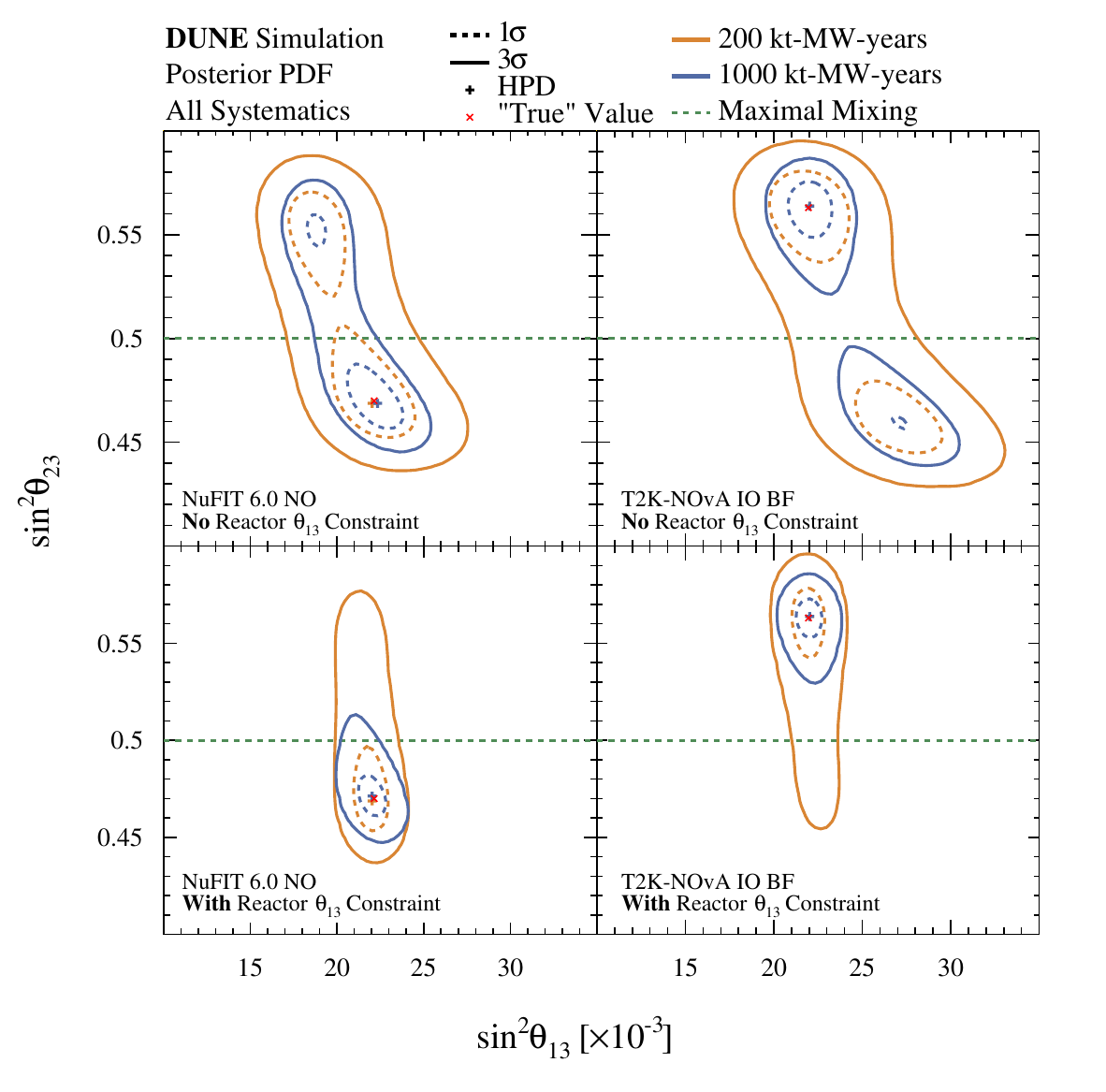}
    \caption{Comparison of the 2D 1$\sigma$ and 3$\sigma$ credible intervals in $\sin^2\theta_{13} - \sin^2\theta_{23}$ with (bottom) and without (top) an external constraint on $\sin^2 \theta_{13}$ from reactor experiments for NuFIT 6.0 (left) and T2K-NOvA (right) Asimov points.}
    \label{fig:Th23Th13}
\end{figure*}

The multidimensional posterior presentation provides a natural way to study the correlations and degeneracies within the parameter space, as well as their evolution with increasing exposure. Information can be lost when considering only isolated one or two-dimensional projections. Given DUNE's goal of making a simultaneous precision measurement of the full three-flavour oscillation model, this Bayesian approach provides a robust framework for characterising the complete parameter space and its correlations.

\subsection{$\theta_{23}$ Octant Sensitivity}

\begin{table}[]
\centering
\renewcommand*{\arraystretch}{1.3}
\begin{tabular*}{\columnwidth}{@{\extracolsep{\fill}}lcccc}
\hline
& \multicolumn{2}{c}{\textbf{No RC}} & \multicolumn{2}{c}{\textbf{With RC}} \\
\textbf{Asimov Value} & \textbf{200} & \textbf{1000} & \textbf{200} & \textbf{1000} \\
\hline
\multicolumn{5}{c}{NuFIT 6.0} \\
$\delta_{CP}=0$       & 1.13 & 3.48 & 4.28 & 208 \\
$\delta_{CP}=-\pi/4$  & 1.11 & 3.02 & 6.09 & 212 \\
$\delta_{CP}=-\pi/2$  & 1.20 & 2.98 & 4.67 & 200 \\
\hline
\multicolumn{5}{c}{T2K--NOvA BF} \\
$\delta_{CP}=0$       & 1.36 & 2.91 & 46.7 & 13100 \\
$\delta_{CP}=-\pi/4$  & 1.46 & 3.51 & 61.4 & 34100 \\
$\delta_{CP}=-\pi/2$  & 1.30 & 6.84 & 58.1 & 19600 \\
\hline
\end{tabular*}
\caption{Bayes factors for the correct $\theta_{23}$ octant, with and without an external constraint on $\sin^2 \theta_{13}$ from reactor experiments (RC) for exposures of 200 and 1000~kt-MW-yr.}
\label{tab:BayesFactors}
\end{table}

\begin{figure*}[t]
    \centering
    \includegraphics[width = \textwidth]{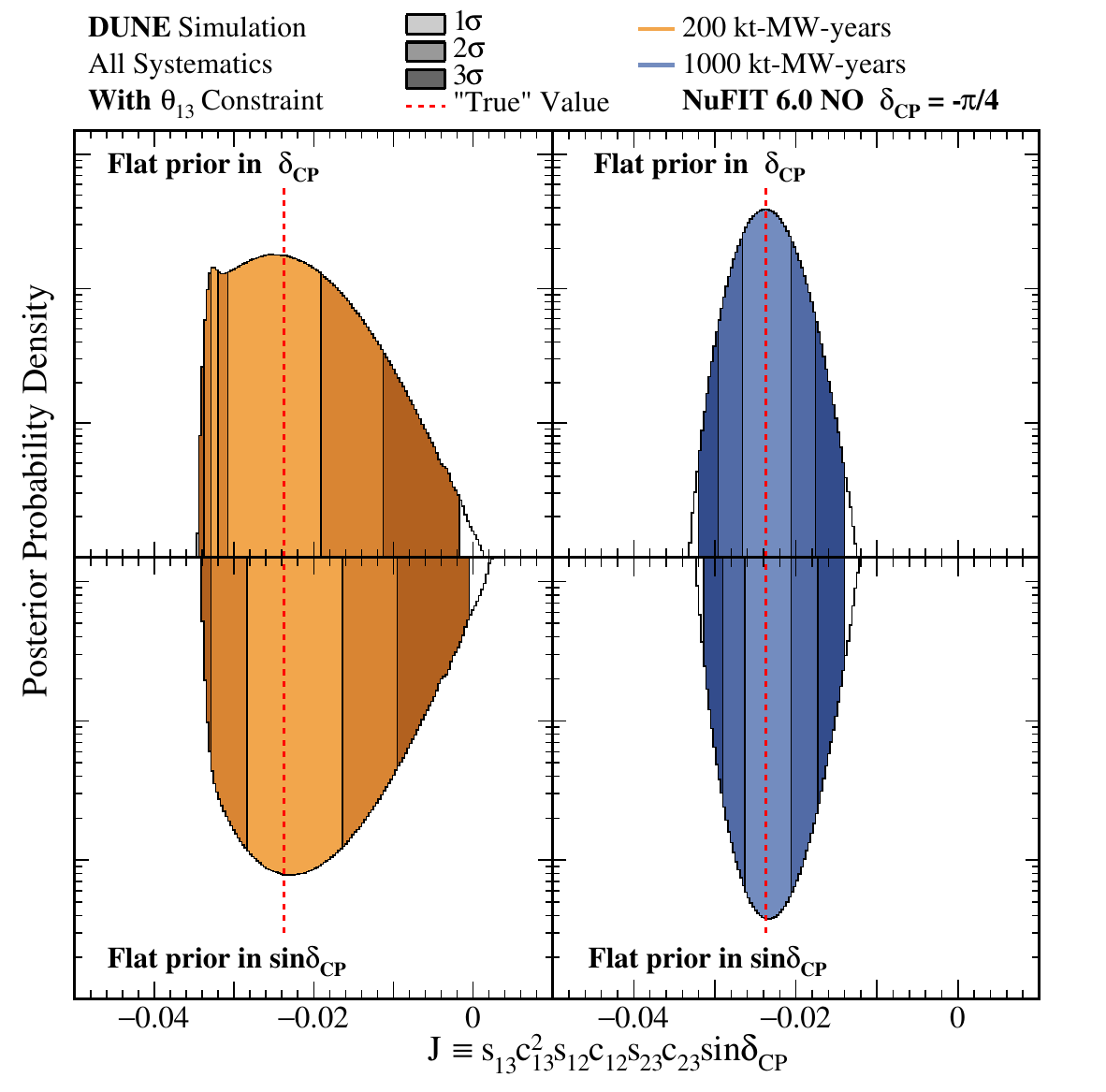}
    \caption{Posterior probability density for the Jarlskog invariant for both 200 (left) and 1000 (right)~kt-MW-yr exposures at the NuFIT 6.0 $\delta_{CP} = -\pi/4$ Asimov point using a uniform prior in $\delta_{CP}$ (top) or in $\sin \delta_{CP}$ (bottom).}
    \label{fig:JCP1D}
\end{figure*}

\begin{figure*}[t]
    \centering
    \includegraphics[width = \textwidth]{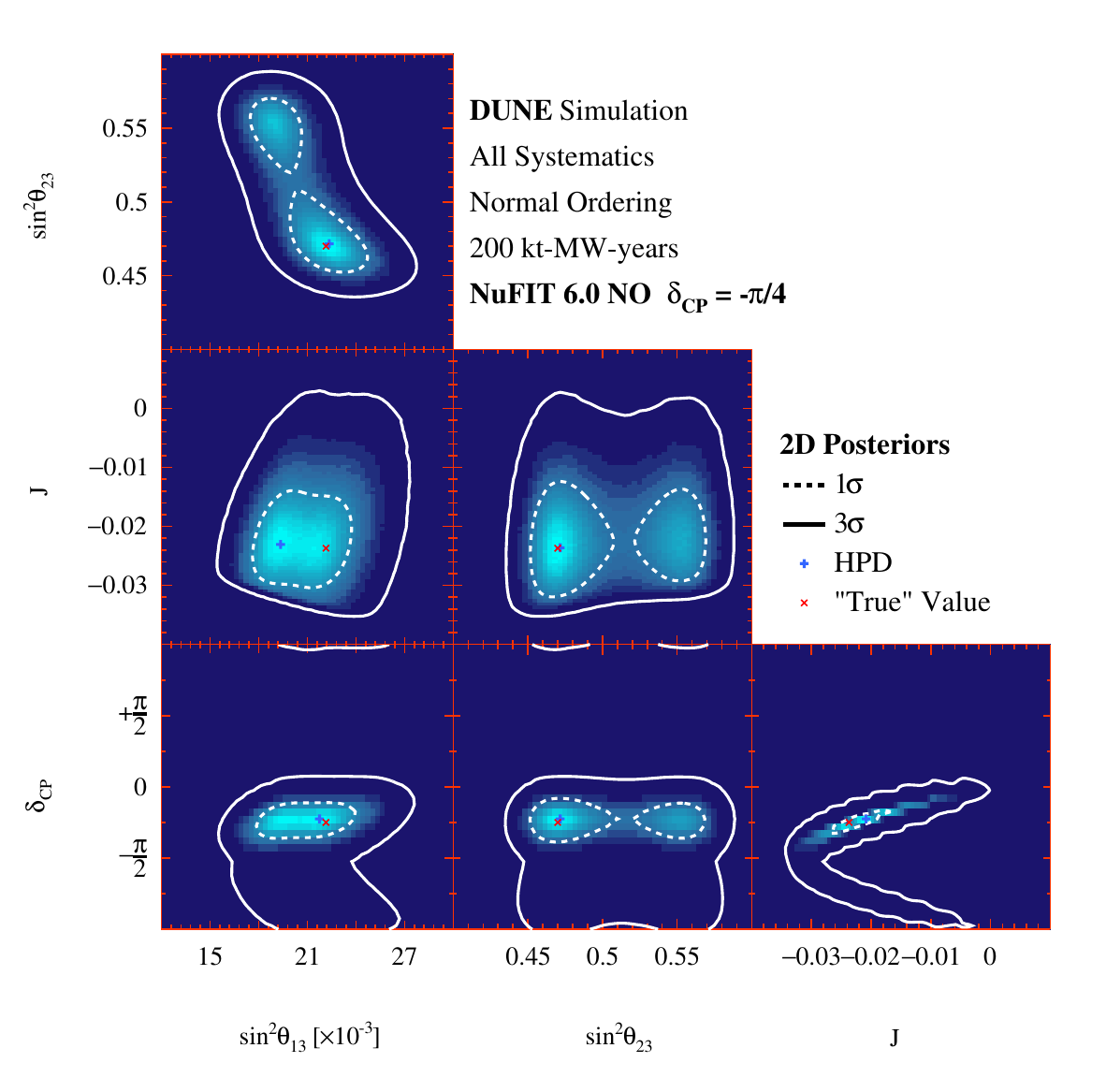}
    \caption{Four-dimensional posterior distribution in the mixing angles and $J$, for 200~kt-MW-yr exposure at NuFIT 6.0 $\delta_{CP} = -\pi/4$ Asimov point using a uniform prior in $\sin \delta_{CP}$. The 2D projections show the 1$\sigma$ and 3$\sigma$ credible intervals.}
    \label{fig:JCP2D}
\end{figure*}

\autoref{fig:Th23Th13} shows a comparison of the 2D credible intervals for $\sin^2\theta_{13} - \sin^2\theta_{23}$ for both NuFIT 6.0 and T2K-NOvA Asimov points, with and without an external constraint on $\sin^2\theta_{13}$ from reactor experiments. The external constraint corresponds to the $\sin^2 \theta_{13}$ central value from the relevant Asimov point and a standard deviation from NuFIT 6.0\footnote{Although NuFIT 6.0 is a global fit to multiple neutrino oscillation experiments, the $\sin^2 \theta_{13}$ constraint is dominated by short-baseline reactor experiments.}. The comparison of sensitivity between the two Asimov points is of interest since global long-baseline neutrino data does not strongly prefer either $\theta_{23}$ octant. There is greater sensitivity to the $\theta_{23}$ octant for the T2K-NOvA Asimov point without the $\sin^2\theta_{13}$ constraint than for the NuFIT 6.0 Asimov point. This is a result of the increased $\nu_{e}$ and $\stackrel{-}{\nu}_e$ events in the upper octant. Maximal mixing does not lie within the 1$\sigma$ contour for the T2K-NOvA Asimov point for the 1000~kt-MW-yr exposure.

In both cases, applying the constraint results in a significant suppression of the posterior in the alternative octant. The "wrong" octant lies outside the 3$\sigma$ credible interval for T2K-NOvA best-fit Asimov point at 1000~kt-MW-yr exposure. This indicates that DUNE can reject the wrong octant at 3$\sigma$ given this true upper octant scenario.

In the Bayesian framework, two models can be compared using the Bayes factor~\cite{Jeffreys, KassRaftery}, which is the ratio of the marginal likelihoods of the two hypotheses. In the case where the priors for both hypotheses are equal, the Bayes factor is equal to the ratio of posterior probabilities. Since a uniform prior is used for $\sin^{2}\theta_{23}$, the Bayes factor for the octant of $\theta_{23}$ can be calculated by taking the ratio of the number steps for $\sin^{2}\theta_{23} > 0.5$ and $\sin^{2}\theta_{23} < 0.5$. \autoref{tab:BayesFactors} shows the Bayes factor for the correct $\theta_{23}$ octant, with and without the $\sin^2\theta_{13}$ constraint, for all Asimov points used in this sensitivity study. As expected, given the posterior distributions shown in this study, at the 200~kt-MW-yr exposure, there is almost no preference for the correct octant at all Asimov points when not including the $\sin^2 \theta_{13}$ constraint. At 1000~kt-MW-yr, the preference for the correct octant improves. There is a similar preference to the octant for both Asimov points when not including the $\sin^2 \theta_{13}$ constraint. When applying the $\sin^2 \theta_{13}$ constraint, there is significantly stronger evidence for the correct octant. The increase is more significant for the T2K-NOvA best-fit Asimov. 

Intuition for the strength of preference indicated by the Bayes factor can be gained by mapping the resulting posterior probabilities to traditional significance thresholds. While posterior probabilities and confidence levels are not directly comparable, the latter are typically derived from Gaussian deviations of a given probability. By analogy, one can associate a Bayes factor with the posterior probability corresponding to a given Gaussian deviation. For example, an equivalent to 3$\sigma$ could be taken to be a Bayes factor of 369.4. In this approach, we find for the T2K-NOvA best-fit Asimov, when including the $\sin^2 \theta_{13}$ constraint, $> 2\sigma$ significance at the 200~kt-MW-yr exposure, and $> 3\sigma$ at the 1000~kt-MW-yr exposure.

The improvement in octant sensitivity obtained from the $\sin^2 \theta_{13}$ constraint illustrates the importance of parameter correlations in oscillation analyses. Within the Bayesian framework, these correlations and their impact on physics conclusions are directly encoded in the posterior distribution. In addition, this approach facilitates the straightforward application of different external constraints, enabling the effect of additional experimental information to be quantified in a computationally efficient manner.

\subsection{Jarlskog Invariant}

In Equation~\ref{eq:Jarls}, a value of $J = 0$ corresponds to CP-conservation, with the degree of CP violation increasing as $\lvert J \rvert$ increases. From the definition, if any of the mixing angle factors is zero, the mixing matrix is CP-conserving. The posterior distribution of $J$ is obtained by taking the MCMC chain and evaluating $J$ at each MCMC step. The simplicity of this procedure is a key advantage of MCMC and Bayesian methods in neutrino oscillation analyses. As mentioned in \autoref{sec:SensMethod}, uniform priors are applied to the oscillation parameters of interest in the form in which they are expressed (e.g. uniform in $\sin^2\theta_{23}$ when fitting $\sin^2\theta_{23}$). This results in a non-uniform prior in $J$ due to the numerous trigonometric dependencies. One might choose to use a uniform prior in $\cos\theta_{23}$ rather than $\sin^2\theta_{23}$ for example when presenting $J$ distributions. Several prior options for mixing angles, not constrained by external data, were studied and found to have no significant effect on the $J$ distribution. 

\autoref{fig:JCP1D} shows the posterior distribution of the Jarlskog invariant for the NuFIT 6.0 $\delta_{CP} = -\pi/4$ Asimov point for exposures of 200 and 1000~kt-MW-yr, using uniform priors in both $\delta_{CP}$ and $\sin\delta_{CP}$. A uniform prior in $\sin\delta_{CP}$ was investigated since $J \propto \sin\delta_{CP}$. As $\delta_{CP}$ is comparatively weakly constrained over the region where the two priors differ most significantly, the choice of prior has a visible effect on the posterior distribution and credible intervals at 200~kt-MW-yr. Nevertheless, the choice of prior does not alter the preference for CP-violating values of $J$. At 1000~kt-MW-yr, however, the posterior becomes largely insensitive to the prior choice, reflecting the increased constraining power of the data. The width of the posterior is significantly reduced at the higher exposure. The region of CP conservation lies just outside the 3$\sigma$ interval at 200~kt-MW-yr, despite the Asimov point corresponding to $\delta_{CP}=-\pi/4$, which lies well away from the point of maximal CP violation.

\autoref{fig:JCP2D} shows the 4D posterior distribution including $J$ and the 3 mixing angles at NuFIT 6.0 $\delta_{CP} = -\pi/4$ Asimov point at a 200~kt-MW-yr exposure, using a uniform priors in $\sin\delta_{CP}$. Although $J$ depends multiplicatively on the mixing angles, the 2D posterior distributions of $J - \sin^2\theta_{13}$ and $J - \sin^2\theta_{23}$ show little correlation. This reflects the fact that DUNE primarily constrains the combined CP-violating amplitude rather than any one parameter in isolation. A stronger relationship can be observed in the 2D posterior of $ \delta_{CP} - J$, highlighting that DUNE will constrain $\delta_{CP}$ directly, leading to a characteristic sine-like relationship between the two quantities.

The Jarlskog invariant provides a clear example of the advantages of the MCMC framework. Although $J$ is not a fit parameter, its posterior distribution, credible intervals and correlations can be obtained without any modification of the underlying fit. This enables a broad range of derived quantities to be studied, allowing DUNE's sensitivity to the full three-flavour oscillation framework to be characterised beyond the parameters explicitly included in the fit.

\section{Conclusion}

This work evaluated DUNE's sensitivity to neutrino oscillation parameters and CPV using a Bayesian statistical approach. The flexibility of the Bayesian framework has been exploited to enable inferences that would otherwise be difficult or computationally expensive. We present the four-dimensional posterior distribution, which naturally shows all correlations between parameters, such as $\sin^2\theta_{23} - \sin^2\theta_{13}$, and highlights DUNE's power to resolve degeneracies at higher exposure. We demonstrate the straightforward incorporation of external constraints within the Bayesian framework by applying an external constraint on $\theta_{13}$, resulting in a significant increase in DUNE's sensitivity to the octant of $\theta_{23}$ for both lower- and upper-octant true values.

We report, for the first time, DUNE's sensitivity to the Jarlskog invariant, $J$, which is impractical to study with the previous frequentist method. This provides a parameterisation-independent measure of CP violation, from which we observe a 3$\sigma$ exclusion of the CP conserving value, $J = 0$, at an exposure of 200~kt-MW-yr for a true $\delta_{CP}$ value of $-\pi/4$. By studying the 2D posterior distributions of $J - \sin^2\theta_{23}$ and $J - \sin^2\theta_{13}$, we show that the appearance spectrum constrains the multiplicative factors entering the Jarlskog invariant, rather than the mixing angles individually.

\begin{acknowledgements}

%
%
%
%
This document was prepared by DUNE collaboration using the resources of the Fermi National Accelerator Laboratory (Fermilab), a U.S. Department of Energy, Office of Science, Office of High Energy Physics HEP User Facility. Fermilab is managed by Fermi Forward Discovery Group, LLC, acting under Contract No. 89243024CSC000002.
%
%
This work was supported by
CNPq,
FAPERJ,
FAPEG, 
FAPESP and,
Funda\c{c}\~{a}o Arauc\'{a}ria,   Brazil;
CFI, 
IPP and 
NSERC,                          Canada;
CERN;
ANID-FONDECYT,                  Chile;
M\v{S}MT,                       Czech Republic;
ERDF, FSE+,
Horizon Europe, 
MSCA and NextGenerationEU,      European Union;
CNRS/IN2P3 and
CEA,                            France;
PRISMA+,                        Germany;
INFN,                           Italy;
FCT,                            Portugal;
CERN-RO/CDI,                        Romania;
NRF,                            South Korea;
Generalitat Valenciana, 
Junta de Andaluc\'{\i}a,
MICINN, and 
Xunta de Galicia,               Spain;
SERI and 
SNSF,                           Switzerland;
T\"UB\.ITAK,                    Turkey;
The Royal Society and 
UKRI/STFC,                      United Kingdom;
DOE and 
NSF,                            United States of America.
%
%
This research used resources of the 
National Energy Research Scientific Computing Center (NERSC), 
a U.S. Department of Energy Office of Science User Facility 
operated under Contract No. DE-AC02-05CH11231.

\end{acknowledgements}

\bibliographystyle{spphys}
\bibliography{ref}

@article{DUNETDR1,
doi = {10.1088/1748-0221/15/08/T08008},
year = {2020},
month = {aug},
publisher = {},
volume = {15},
number = {08},
pages = {T08008},
author = {B. Abi and others},
title = {Volume I. Introduction to DUNE},
journal = {Journal of Instrumentation}
}

@misc{DUNETDR2,
      title={Deep Underground Neutrino Experiment (DUNE), Far Detector Technical Design Report, Volume II: DUNE Physics}, 
      author={B. Abi and others},
      year={2020},
      eprint={2002.03005},
      archivePrefix={arXiv},
      primaryClass={hep-ex}, 
}

@article{leptonquark,
title = {Neutrino mass hierarchy},
journal = {Progress in Particle and Nuclear Physics},
volume = {83},
pages = {1-30},
year = {2015},
issn = {0146-6410},
doi = {10.1016/j.ppnp.2015.05.002},
author = {X. Qian and P. Vogel}
}

@article{fukugita,
title = {Barygenesis without grand unification},
journal = {Physics Letters B},
volume = {174},
number = {1},
pages = {45-47},
year = {1986},
issn = {0370-2693},
doi = {10.1016/0370-2693(86)91126-3},
author = {M. Fukugita and T. Yanagida}
}

@article{davidson,
title = {Leptogenesis},
journal = {Physics Reports},
volume = {466},
number = {4},
pages = {105-177},
year = {2008},
issn = {0370-1573},
doi = {10.1016/j.physrep.2008.06.002},
author = {Sacha Davidson and Enrico Nardi and Yosef Nir}
}

@article{mr2t2,
 ISSN = {00063444},
 author = {W. K. Hastings},
 journal = {Biometrika},
 number = {1},
 pages = {97--109},
 publisher = {[Oxford University Press, Biometrika Trust]},
 title = {Monte Carlo Sampling Methods Using Markov Chains and Their Applications},
 volume = {57},
 year = {1970}
}

@article{amcmc,
  author  = {H. Haario and E. Saksman and J. Tamminen},
  title   = {An adaptive Metropolis algorithm},
  journal = {Bernoulli},
  volume  = {7},
  number  = {2},
  pages   = {223--242},
  year    = {2001}
}

@article{neutrinotheory,
title = {CP violation and neutrino oscillations},
journal = {Progress in Particle and Nuclear Physics},
volume = {60},
number = {2},
pages = {338-402},
year = {2008},
issn = {0146-6410},
doi = {10.1016/j.ppnp.2007.10.001},
author = {Hiroshi Nunokawa and Stephen Parke and José W.F. Valle}
}

@article{DUNESens,
  author  = {B. Abi and others},
  title   = {Long-baseline neutrino oscillation physics potential of the DUNE experiment},
  journal = {European Physical Journal C},
  volume  = {80},
  number  = {10},
  pages   = {978},
  year    = {2020},
  doi     = {10.1140/epjc/s10052-020-08456-z}
}

@article{numi,
  title = {Neutrino flux predictions for the NuMI beam},
  author = {Aliaga, L. and others},
  collaboration = {MINER\ensuremath{\nu}A Collaboration},
  journal = {Phys. Rev. D},
  volume = {94},
  issue = {9},
  pages = {092005},
  numpages = {10},
  year = {2016},
  month = {Nov},
  publisher = {American Physical Society},
  doi = {10.1103/PhysRevD.94.092005}
}

@article{CVN,
  title = {Neutrino interaction classification with a convolutional neural network in the DUNE far detector},
  author = {B. Abi and others},
  collaboration = {DUNE Collaboration},
  journal = {Phys. Rev. D},
  volume = {102},
  issue = {9},
  pages = {092003},
  numpages = {20},
  year = {2020},
  month = {Nov},
  publisher = {American Physical Society},
  doi = {10.1103/PhysRevD.102.092003}
}

@article{asimov,
  author       = {Glen Cowan and Kyle Cranmer and Eilam Gross and Ofer Vitells},
  title        = {Asymptotic formulae for likelihood-based tests of new physics},
  journal      = {European Physical Journal C},
  volume       = {71},
  number       = {1554},
  year         = {2011},
  pages        = {1554},
  doi          = {10.1140/epjc/s10052-011-1554-0}
}

@article{nufit6,
   title={NuFit-6.0: updated global analysis of three-flavor neutrino oscillations},
   volume={2024},
   ISSN={1029-8479},
   DOI={10.1007/jhep12(2024)216},
   number={12},
   journal={Journal of High Energy Physics},
   publisher={Springer Science and Business Media LLC},
   author={Esteban, Ivan and Gonzalez-Garcia, M. C. and Maltoni, Michele and Martinez-Soler, Ivan and Pinheiro, João Paulo and Schwetz, Thomas},
   year={2024},
   month=dec }

@article{t2knova,
	author = {{The NOvA Collaboration and The T2K Collaboration}},
	date = {2025/10/01},
	doi = {10.1038/s41586-025-09599-3},
	id = {Abubakar2025},
	isbn = {1476-4687},
	journal = {Nature},
	number = {8086},
	pages = {818--824},
	title = {Joint neutrino oscillation analysis from the T2K and NOvA experiments},
	volume = {646},
	year = {2025}
}

@book{Jeffreys,
  author    = {Harold Jeffreys},
  title     = {Theory of Probability},
  edition   = {3},
  year      = {1961},
  publisher = {Oxford University Press},
  address   = {Oxford, England}
}

@article{KassRaftery,
  author  = {Robert E. Kass and Adrian E. Raftery},
  title   = {Bayes Factors},
  journal = {Journal of the American Statistical Association},
  volume  = {90},
  number  = {430},
  pages   = {773--795},
  year    = {1995},
  doi     = {10.1080/01621459.1995.10476572}
}

@article{CUDAProb3,
title = {Massively parallel computation of atmospheric neutrino oscillations on CUDA-enabled accelerators},
journal = {Computer Physics Communications},
volume = {234},
pages = {235-244},
year = {2019},
issn = {0010-4655},
doi = {10.1016/j.cpc.2018.07.022},
author = {Felix Kallenborn and Christian Hundt and Sebastian Böser and Bertil Schmidt}
}

@article{GENIEManual,
    author = "Andreopoulos, Costas and Barry, Christopher and Dytman, Steve and Gallagher, Hugh and Golan, Tomasz and Hatcher, Robert and Perdue, Gabriel and Yarba, Julia",
    title = "{The GENIE Neutrino Monte Carlo Generator: Physics and User Manual}",
    eprint = "1510.05494",
    archivePrefix = "arXiv",
    primaryClass = "hep-ph",
    reportNumber = "FERMILAB-FN-1004-CD",
    month = "10",
    year = "2015"
}

@article{GENIE,
    author = "Andreopoulos, C. and others",
    title = "{The GENIE Neutrino Monte Carlo Generator}",
    eprint = "0905.2517",
    archivePrefix = "arXiv",
    primaryClass = "hep-ph",
    reportNumber = "FERMILAB-PUB-09-418-CD",
    doi = "10.1016/j.nima.2009.12.009",
    journal = "Nucl. Instrum. Meth. A",
    volume = "614",
    pages = "87--104",
    year = "2010"
}

@misc{MaCh3,
  author       = {{The MaCh3 Collaboration}},
  title        = {{mach3-software/MaCh3: v2.0.0}},
  year         = "2025",
  publisher    = {Zenodo},
  version      = {v2.0.0},
  doi          = {10.5281/zenodo.15413080},
  url          = {https://doi.org/10.5281/zenodo.15413080}
}

@article{PDGNeutrinoMixing,
    author = "Takahashi, F. and others",
    collaboration = "Particle Data Group",
    title = "{Review of Particle Physics}",
    doi = "10.1142/S0217751X26300115",
    journal = "Int. J. Mod. Phys. A",
    volume = "41",
    pages = "2630011",
    year = "2026"
}

@misc{cafana,
  author       = {{NOvA Collaboration}},
  title        = {{NOvA-ART, Chapter: CAFAna Overview}},
  year         = {2019},
  howpublished = {\url{https://cdcvs.fnal.gov/redmine/projects/novaart/wiki/CAFAna_overview}},
  note         = {Edited by the NOvA Collaboration}
}

@article{Jarlskog,
  author  = {Jarlskog, Cecilia},
  title   = {Commutator of the Quark Mass Matrices in the Standard Electroweak Model and a Measure of Maximal CP Violation},
  journal = {Phys. Rev. Lett.},
  volume
  = {55},
  pages   = {1039},
  year    = {1985},
  doi     = {10.1103/PhysRevLett.55.1039}
}

@article{LowExposure,
  author        = {Abud, A. A. and others},
  collaboration = {DUNE Collaboration},
  title         = {Low Exposure Long-Baseline Neutrino Oscillation Sensitivity of the DUNE Experiment},
  journal       = {Physical Review D},
  volume        = {105},
  number        = {7},
  pages         = {072006},
  year          = {2022},
  doi           = {10.1103/PhysRevD.105.072006},
  eprint        = {2109.01304},
  archivePrefix = {arXiv},
  primaryClass  = {hep-ex}
}

@book{PriorReweight,
  author    = {Christian P. Robert and George Casella},
  title     = {Monte Carlo Statistical Methods},
  edition   = {2},
  year      = {2004},
  publisher = {Springer},
  address   = {New York},
  series    = {Springer Texts in Statistics},
  isbn      = {978-0-387-21239-5},
  doi       = {10.1007/978-1-4757-4145-2}
}

@article{NueScatter,
  author        = {Marshall, C. M. and McFarland, K. S. and Wilkinson, C.},
  title         = {Neutrino-electron elastic scattering for flux determination at the DUNE oscillation experiment},
  journal       = {Phys. Rev. D},
  volume        = {101},
  number        = {3},
  pages         = {032002},
  year          = {2020},
  doi           = {10.1103/PhysRevD.101.032002},
  eprint        = {1910.10996},
  archivePrefix = {arXiv},
  primaryClass  = {hep-ex}
}

@article{PandoraSoft,
   title={The Pandora software development kit for pattern recognition},
   volume={75},
   ISSN={1434-6052},
   DOI={10.1140/epjc/s10052-015-3659-3},
   number={9},
   journal={The European Physical Journal C},
   publisher={Springer Science and Business Media LLC},
   author={Marshall, J. S. and Thomson, M. A.},
   year={2015},
   month={Sept} 
}

@article{PandoraDUNE,
   title={Neutrino interaction vertex reconstruction in DUNE with Pandora deep learning},
   volume={85},
   ISSN={1434-6052},
   DOI={10.1140/epjc/s10052-025-14313-8},
   number={6},
   journal={The European Physical Journal C},
   publisher={Springer Science and Business Media LLC},
   author={Abud, A. Abed and others},
   year={2025},
   month={June} }

\end{document}